\documentclass[reprint,prb,twocolumn,showpacs,superscriptaddress,aps]{revtex4-1}

\usepackage{graphicx}
\DeclareGraphicsExtensions{.png,.jpg,.eps}

\usepackage{xcolor}
\usepackage{amsmath}

\begin{document}

\title{Dirac topological insulator in the d$_{z^2}$ manifold of a honeycomb oxide}

\author{J. L. Lado}
\affiliation{International Iberian Nanotechnology Laboratory, Braga, Portugal}  
  \email{jose.luis.lado@gmail.com}
	
\author{V. Pardo}
\affiliation{Departamento de F\'{i}sica Aplicada,
  Universidade de Santiago de Compostela, E-15782 Campus Sur s/n,
  Santiago de Compostela, Spain}
\affiliation{Instituto de Investigaci\'{o}ns Tecnol\'{o}xicas,
  Universidade de Santiago de Compostela, E-15782 Campus Sur s/n,
  Santiago de Compostela, Spain}  
  \email{victor.pardo@usc.es}

\date{\today} 

\begin{abstract} 


We show by means of ab initio calculations and tight-binding modeling that an
oxide system based on a honeycomb lattice can sustain topologically non-trivial
states if a single orbital dominates the spectrum close to the Fermi level. 
In such situation, the low energy spectra is described by two Dirac
equations that become non-trivially gapped
when spin-orbit coupling (SOC) is switched on. 
We provide one specific example for this but the recipe is general.
We
discuss a realization of this starting from a conventional spin-a-half
honeycomb antiferromagnet whose states close to the Fermi energy
are d$_{z^2}$ orbitals. Switching off magnetism by atomic substitution
and ensuring that the electronic structure
becomes two-dimensional is sufficient for topologicality to arise in such a
system. 
We show that the gap in such model scales linearly
with SOC, opposed to
other oxide-based topological insulators, where smaller gaps tend to appear by
construction of the lattice. We also
provide a study of the quantum Hall
effect in such system, showing the close connections with the physics of
graphene but in a d-electron system.

\end{abstract}
\maketitle

\section{Introduction}

Topological insulators\cite{hasan2010colloquium,qi2011topological} 
(TI) are a new class of materials that present a gap in the
bulk but on their edges they have gapless states that conduct without
dissipation. They have produced a renewed understanding of
the topological properties of the single-particle Hamiltonian that describes
the electronic structure of an insulator. 
The topological properties
of the edge states arise due to a finite Chern number as in the quantum Hall
effect,\cite{hatsugai1993chern,thouless1982quantized}
they can be protected by different symmetries of the Hamiltonian,
such as crystalline \cite{fu2011topological}
or time reversal symmetry,\cite{kane2005z}
or they can be arise due to
 interaction effects in topological
Mott insulators\cite{tmi_balents}
and topological Kondo insulators.\cite{dzero2010topological}
Among the different possible classes, TI with time reversal (TR) symmetry\cite{hasan2010colloquium}
show spin-momentum
locking in their edge states.
All sorts of peculiar physics has been predicted to arise from them, such as
topological superconductors showing\cite{qi2011topological} Majorana
bound states\cite{ti_majo} with anyonic
statistics.\cite{hastings2013metaplectic}
In terms of applications, those
dissipation-less edge state could help
in transmitting information
without losses,\cite{jiang2009topological}
find applications in spintronic devices\cite{pesin2012spintronics}
or become a key ingredient for topological
quantum computing.\cite{ti_majo,laflamme2014hybrid}

Topological insulators can appear in two or three dimensional systems. In two-dimensions, the first proposal for a topological insulator was the
prediction of a quantum spin Hall effect,\cite{ti_1} a peculiar type of quantum Hall
effect that is not caused by an external magnetic field but by the internal
spin-orbit interaction of the material. This was first proposed to appear in
graphene\cite{ti_2} but not materialized experimentally. The same
ideas were shown to operate in HgTe/CdTe quantum
wells,\cite{hgte_cdte_prediction} where a larger SOC strength
takes place and allowed for its experimental observation.\cite{hgte_cdte_molenkamp}

Oxides provide also an interesting platform for non-trivial topological
properties to appear. Early since the discovery of topological insulators,
several oxides with different structures have been identified as possible
topological insulators: honeycomb-based iridates,\cite{na2iro3}
pyrochlores,\cite{pyrochlores_ti} corundum structure\cite{corundum,afonso}
and rutiles.\cite{vpardo_Dirac,cro2_nano}
The honeycomb lattice has been the simplest to investigate. Based on these concepts, Xiao et al.\cite{xiao} proposed to build an oxide-based honeycomb
lattice, though largely buckled, using perovskite bilayers grown along the
perovskite (111) direction, and other authors have followed suit, showing that
this kind of phases appear ubiquitously, even for the low SOC
limit.\cite{pentcheva_111} It was shown\cite{lado} that the
size of the topological gap depends on the
strength of the trigonal splitting, and only inversely with SOC
strength, hence these phases show up also for 3d electron systems, but there
they compete with Jahn-Teller distortions, charge and magnetic ordering.
\cite{xiao,lado,afonso}

Here we propose a different oxide-based route to realize topologically non-trivial properties. The system under study shows a
low-energy electronic structure
identical to the one of graphene, but with $d_{z^2}$
orbitals dominating the spectrum around the Fermi level instead of the $p_z$ states of graphene.
Our proposal is based of atomic substitution
on the recently synthesized compound
InCu$_{2/3}$V$_{1/3}$O$_3$,\cite{incuv,yan2012magnetic,yehia2010finite} which shows a honeycomb
lattice of S=1/2 Cu$^{2+}$ cations. We propose to substitute Cu by
isoelectronic Ag\cite{bratsch2014aag2} 
and isolate the active Ag-rich planes by a Zn-rich layer to
produce a truly two-dimensional electronic structure.
In InZn$_{1/3}$Ag$_{1/3}$V$_{1/3}$O$_3$,
due to the interplay of local
$C_3$ symmetry, crystal field and charge transfer,
close to the Fermi energy
the band structure is composed
by Ag $d_{z^2}$  orbitals
forming an effective honeycomb lattice.
In the absence of SOC, the low-energy spectrum
is described by two gapless Dirac equations close to the $K$ and $K'$
points.
When relativistic
effects are included, the system opens up a 
gap and shows a topological invariant $\nu=-1$, realizing
the quantum spin Hall effect as a d$_{z^2}$ version
of the Kane-Mele model.\cite{ti_2}
We
show that the quantum spin Hall effect
is resilient to certain time reversal symmetry breaking, such
as off-plane magnetism and gauge magnetic fields.
We also show that the gauge magnetic field yields
a ferromagnetic quantum spin Hall state, showing
the Landau level structure typical of Dirac fermions.
Finally, we summarize our conclusions.

\section{Electronic structure analysis}

\subsection{Antiferromagnetic Cu based compound}

\begin{figure}[t!]
\begin{center}
\includegraphics[width=\columnwidth]{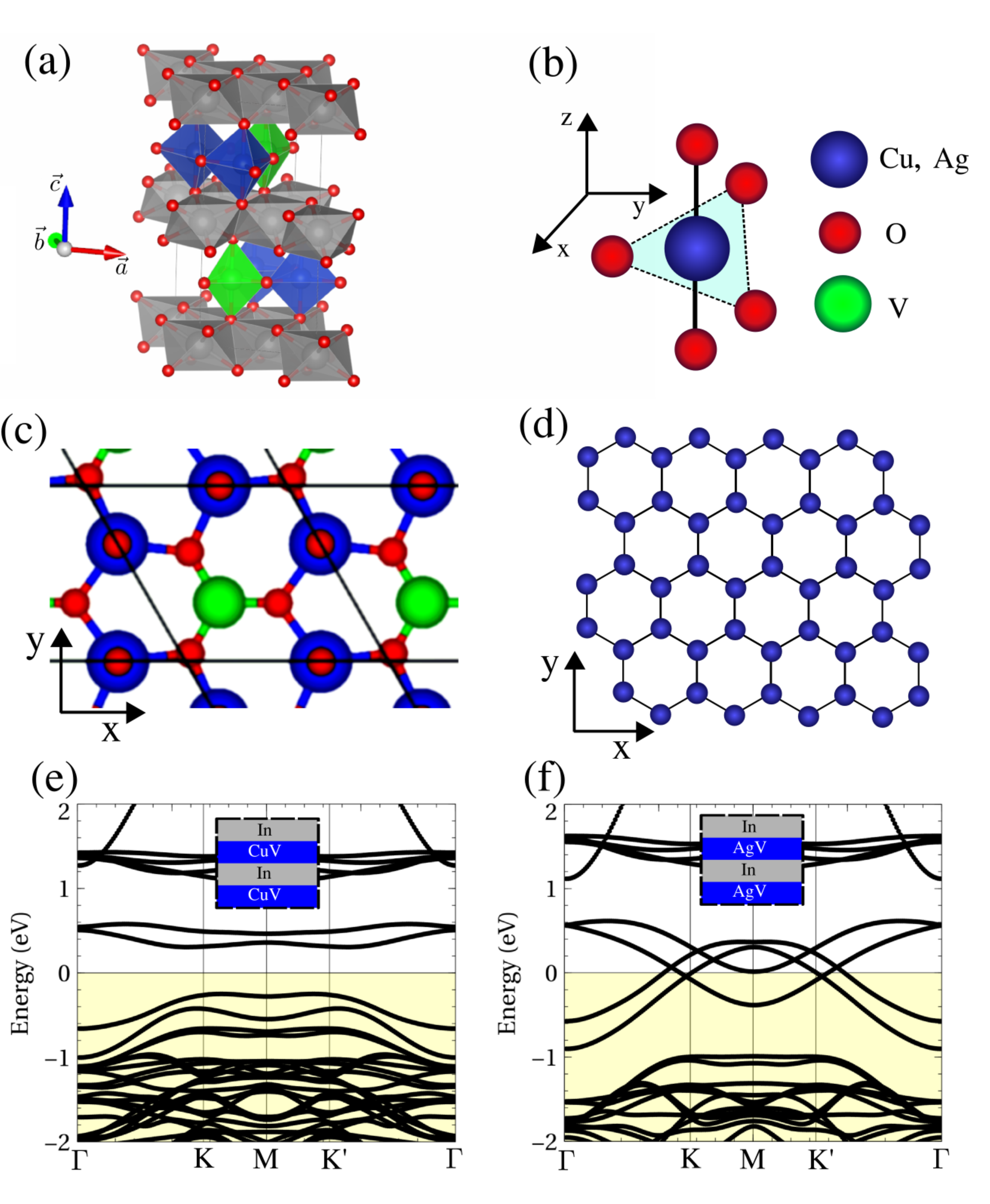}
\caption{(Color online.)
(a) Unit cell of the crystal,
showing alternating planes of Cu(Ag),V and In. 
Panel (b) shows a sketch of the Cu(Ag) oxygen environment, 
giving rise
to local $C_3$ symmetry. Panel (c) shows the Ag,V plane,
where the Ag atoms
form a (slightly distorted) honeycomb lattice
as shown in (d).
Panel (e) shows the
band structure of InCu$_{2/3}$V$_{1/3}$O$_3$ calculated at the GGA
level, yielding a gapped antiferromagnetic ground state.
In comparison,
for InAg$_{2/3}$V$_{1/3}$O$_3$ 
the ground state is non magnetic as
shown in (f).
In both cases,
the bands just around the Fermi level are d$_{z^2}$ from Cu for (e)
and Ag for (f).
}
\label{bs_ag}
\end{center}
\end{figure}

We start describing the electronic structure
of the recently synthesized compound
InCu$_{2/3}$V$_{1/3}$O$_3$,\cite{incuv} which is formed by a honeycomb
lattice of S=1/2 Cu$^{2+}$ cations developing
an antiferromagnetic order.  Even though its structure
(see Fig. \ref{bs_ag}a) looks very
complicated, composed by alternating layers
of In and Cu,V within an oxygen environment,
its electronic structure turns out to be very simple. With a
simple ionic model one can do the following electron count: In$^{3+}$,
V$^{5+}$: d$^0$ (both full/empty shell ions) and Cu$^{2+}$: d$^9$ (S=1/2).
The local environment of Cu is CuO$_5$, with the
z-axis being a C$_3$ symmetry axis and the three in-plane oxygens forming
120$^{\circ}$ (see Fig. \ref{bs_ag}b,c). In this situation the
d$_{z^2}$ orbitals lie
higher in energy, half filled, whereas the other d-orbitals
are completely filled.
Therefore, close to the Fermi level, only the higher-lying
d$_{z^2}$ orbitals are present, substantially separated
and electronically decoupled from all other
d orbitals. This
gives rise
to an effective single orbital weakly coupled pair of honeycomb lattices,
one per each Cu plane in the unit cell
(Fig. \ref{bs_ag}d).
This decoupling of the $d_{z^2}$ bands can be understood by the
local $C_3$ symmetry, which allows to expand the polar dependence
of the local potential as
$V_{C_3}(r,\theta,\phi) = \sum_n a_n e^{i3\phi n} v_n(r,\theta)$, with
$\phi$ the polar angle. 
The previous expansion guarantees that
locally there is no mixing between $d_{z^2} (m=0)$ and the other 
d-orbitals
 $(m =\pm 1$ i.e. the xz/yz orbitals, $\pm 2$ i.e. the xy/$x^2-y^2$ ones),
given that any matrix element of the form
$\langle m=0 | e^{i3\phi n} | m=\pm1,\pm2 \rangle$
is identically zero for $n \neq 0$.
This can be seen in Fig. \ref{bs_ag}e, where the band structure of
the antiferromagnetic Cu-based
system is presented and the bands both just above and below the Fermi level
have Cu $d_{z^2}$ symmetry. Thus, to some extent, such a complicated material
like
InCu$_{2/3}$V$_{1/3}$O$_3$ turns out to be effectively a single-band compound.
Opposite to high-T$_c$ cuprates, where the d$_{x^2-y^2}$ band occurs around the Fermi
level, here it is the d$_{z^2}$ band that is central. The nearest-neighbor
antiferromagnetic coupling produces a 2D-AF, which is the ground state even
without correlations due to the low bandwidth of the Cu $d_{z^2}$ bands. Experimentally, a large J$_{AF}$= 120 K has been estimated, with a magnetic phase transition of some sort occurring at 38 K.\cite{incuv}

\subsection{Non magnetic Ag based compound}

A simple way to remove magnetism in this system is to
substitute directly Cu by isoelectronic Ag. Doing this, we change from 
a 3d to a  4d electron system, with the corresponding increase in spin-orbit
coupling strength and band width.
The solution
obtained for the compound InAg$_{2/3}$V$_{1/3}$O$_3$ is non-magnetic and its
band structure is shown in Fig. \ref{bs_ag}f.
The electron count is basically
retained: V$^{5+}$:d$^0$, empty-shell nonmagnetic cations and Ag$^{2+}$:d$^9$
atoms with a single hole in the aforementioned d$_{z^2}$ orbital, that becomes
decoupled by the peculiar AgO$_5$ environment this structure presents. In
this situation, the crystal-field energy that separates the d$_{z^2}$ bands
from the rest is even higher due to the more delocalized nature of the Ag 4d
electrons, and hence this system becomes even more one-orbital-like. We can see
that there is a 1.5 eV energy window around the Fermi level completely dominated by the four Ag
d$_{z^2}$ bands in the unit cell. 
However the two Ag-rich planes
show a sizable interaction
due to their proximity, so that
the low-energy properties
are not the ones of two decoupled
honeycomb lattices.

\subsection{Dirac-like Ag-Zn based compound}

\begin{figure}[t!]
\begin{center}
\includegraphics[width=\columnwidth]{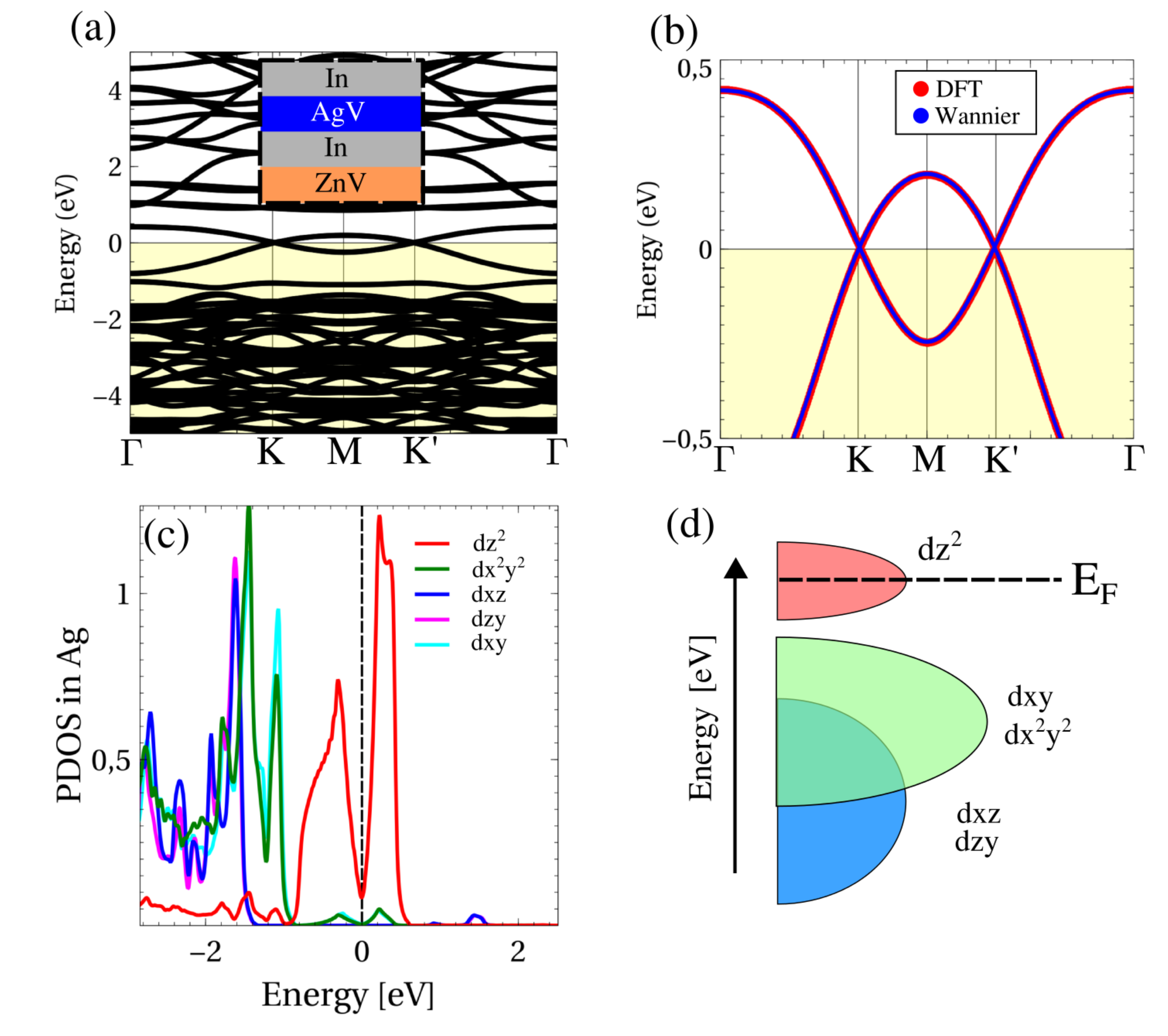}
\caption{(Color online.) 
(a) Non relativistic band structure of
InZn$_{1/3}$Ag$_{1/3}$V$_{1/3}$O$_3$ calculated at the GGA level, where
the low-energy properties are dominated by the single Ag layer.
We observe that
the Dirac point occurs close to the K point. Panel (b)
shows a comparison of the low-energy bands between DFT and the
tight binding model obtained through Wannierization. 
Although the electronic spectrum
is Dirac-like, there is a strong electron-hole asymmetry.
Projected density of states (c) over the Ag atom shows
the dominant d$_{z^2}$ character of the Dirac bands, that
can be understood with the simplified picture (d) considering the three oxygens
in-plane create a local 120$^{\circ}$ rotation symmetry.
}\label{bs_zn}
\label{mono_ag}
\end{center}
\end{figure}

In order to produce
a single Dirac point in the band structure and change the topological
properties of the system, one can try to remove this
interaction between the active Ag layers 
by
adding a spacing layer between them. Ideally, the
preferred experimental situation would be to grow a single layer of this
honeycomb lattice on the appropriate substrate (typically Al$_2$O$_3$ or some
other hexagonal-based substrate like a wurtzite nitride AlN, e.g.).
Another approach is to
introduce a spacing layer based on full-shell Zn$^{2+}$:d$^{10}$ cations in
the place
of Ag (Figs. \ref{mono_ag}a,b). 
That way, interactions between the active Ag orbitals along the z-axis
will be largely reduced 
and the problem becomes purely two-dimensional.
Calculation of the orbital resolved density of states in Ag
(Fig. \ref{mono_ag}c)
confirms that in such situation the low-energy levels
are still Ag d$_{z^2}$, showing also
a strong mixing between the other d-orbitals away from the Fermi level (Fig. \ref{mono_ag}d).

In the situation described above, we
can observe in Fig. \ref{mono_ag}a the band structure of
that layered alternation of Zn-rich and Ag-rich planes in the two honeycombs
that form the unit cell. Due to
the lack of interaction between active Ag-rich planes, 
the low-energy spectrum are two Dirac cones
close to the K and K' points, 
as expected from a simple one-orbital picture on a purely
two-dimensional honeycomb
lattice. Such electronic structure (see Fig. \ref{mono_ag}b) can be easily captured
by means of Wannier procedure projecting
onto the Ag d-orbitals, which gives rise to a
tight binding Hamiltonian of the form

\begin{equation}
H_W = \sum_{i,j} t_{ij} c_{j}^\dagger c_{i}
\end{equation}
where $c_i$ and $c^\dagger_j$ are (spinless) annihilation and creation operators
in the d-like Wannier Ag orbitals, and $t_{i,j}$ are the
hopping parameters obtained with the Wannierization procedure.

It is worth to mention that
the underlying Ag honeycomb lattice
has a small distortion
due to a small dimerization
between pairs of Ag sites.
The effect
of such structural feature is that the Dirac points are
not strictly located in the K and K' points, but
have a very small displacement in
reciprocal space.\cite{pereira2009tight,oliva2016effective}

\section{Topological insulating state in Ag-Zn multilayer}

The calculations presented so far do not consider
relativistic effects, so that the low-energy
Hamiltonian of the Ag-Zn compound are two gapless Dirac equations.
In comparison, 
when SOC is included in the ab initio calculations, 
we observe that a gap opens up in the Dirac points (Fig. \ref{mono_soc}a).
The former situation is analogous to the behavior of
other Dirac materials as graphene, where an opening of the Dirac points
by a perturbation which does not break time reversal
symmetry nor inversion leads to a topological insulator
of the $Z_2$ topological class.\cite{ti_2}

\subsection{Topological invariant}

In crystals with inversion symmetry,
the non-triviality of the band structure
can be easily checked by calculating the
parities of the Khon-Sham eigenvalues.\cite{fu_kane}
However, the unit cell of 
the Ag-Zn compound no longer preserves inversion symmetry,
invalidating the previous approach.
To check that the band structure of this system
is topologically non-trivial, we will instead
take advantage of the tight binding Hamiltonian
obtained previously.
We generate a relativistic
Hamiltonian by taking the Wannier Hamiltonian 
and adding SOC
as an atomic term of the form
$H_{SOC} = \lambda_{SOC}\vec L \cdot \vec S$. 
This leads to the following
relativistic tight binding Hamiltonian

\begin{multline}
H_0 = H_{W} + H_{SOC} = \\ = \sum_{i,j,s} t_{ij} c_{j,s}^\dagger c_{i,s} 
+ \sum_{i,j,s,s'}\lambda_{SOC} \vec L_{i,j} \cdot \vec S_{s,s'} c^\dagger_{j,s} c_{i,s'}
\end{multline}
where $t_{ij}$ are the (spinless) hopping parameters
obtained from the Wannierization technique, 
$c_{i,s}$ and $c^\dagger_{j,s'}$ spinful annihilation and creation
operators,
$\vec L$ the orbital
angular momentum in the d-orbitals and $\vec S$ the spin Pauli matrices.
This approach allows to artificially control the strength of the
SOC without having to perform further ab initio calculations.
This is specially useful to understand the mechanism of gap opening
as we will see later.

\begin{figure}[t!]
\begin{center}
\includegraphics[width=\columnwidth]{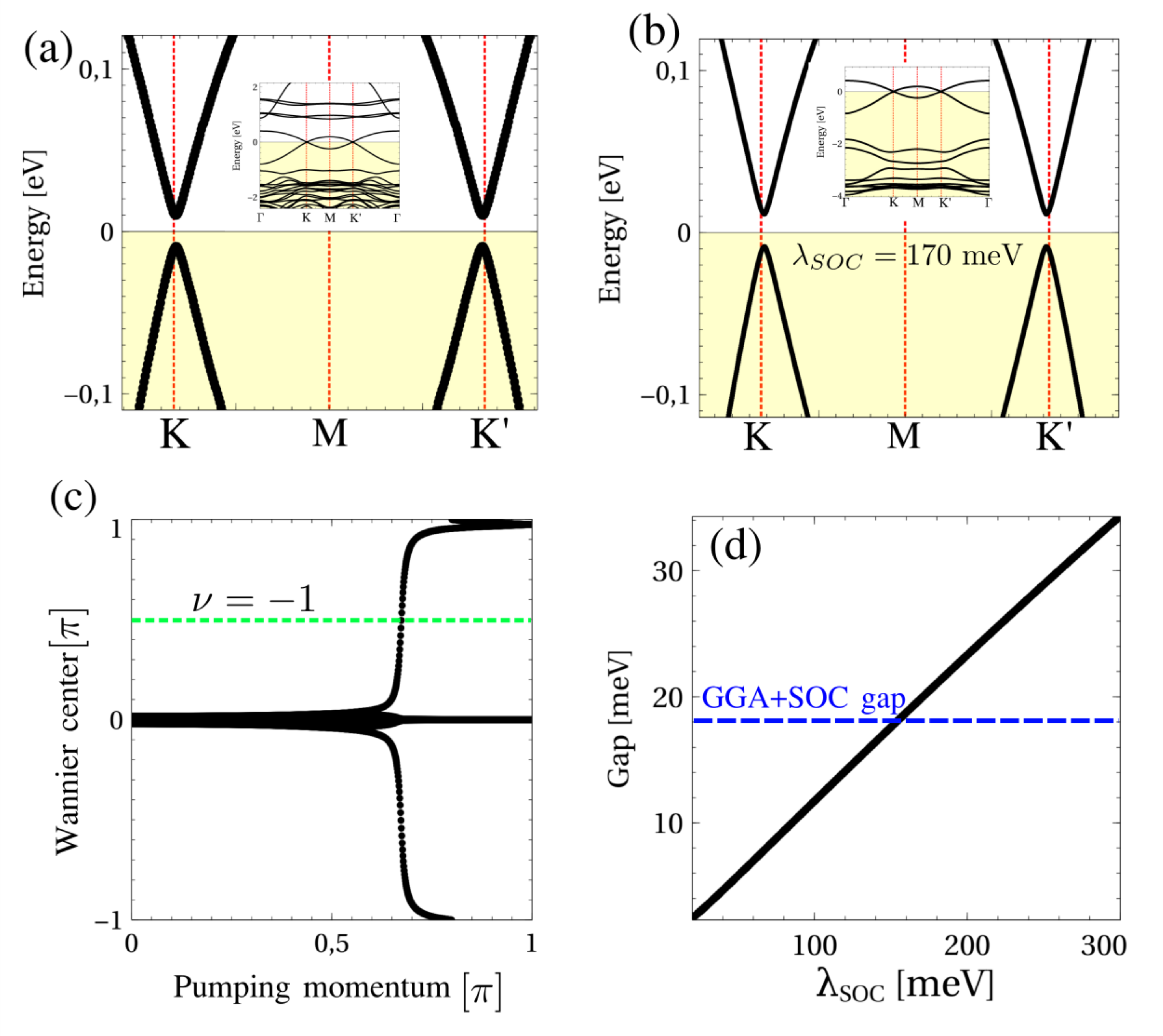}
\caption{(Color online.)
Ab initio band structure (a) with spin-orbit coupling introduced in a fully relativistic manner by using the {\sc elk} code, showing
a band gap opening close to the Dirac points. Panel (b) shows the band
structure calculated using the tight-binding model obtained through Wannierization plus atomic SOC, which agrees very well
with the one obtained from DFT. With the tight binding model, the flow of the
Wannier charge centers, black dots in (c), is calculated as a function of a
the other crystal momentum, which is used as pumping parameter.
The fact that the Wannier centers cross the dashed green line an odd number
of times proves
the topological non-trivial nature of
the band gap. 
Panel (d) shows the evolution of the gap with the SOC strength,
showing a linear dependence and allowing for a determination of $\lambda_{SOC}$ by comparing with the ab initio calculation.
}
\label{model}
\label{mono_soc}
\end{center}
\end{figure}

Upon introduction of SOC in the Wannier Hamiltonian
we obtain a band structure showing a band gap (Fig. \ref{model}b), 
in agreement with the
fully relativistic ab initio calculations (Fig. \ref{model}a). 
With the tight binding model,
the non-triviality of the band gap can be probed
by calculating the $Z_2$ invariant by means
of the flow of Wannier charge centers.\cite{soluyanov2011computing} 
Figure \ref{model}c confirms that the system
is non-trivial because an odd number of crossings occurs along the variation of the chosen cyclic parameter, in this case the crystal momentum.

An interesting property of the gap of this system is that it depends
linearly with the SOC strength (Fig. \ref{model}d), 
meaning that the channel that opens up
the band gap enters as first order in 
SOC.\cite{gmitra2009band}
This is an important
finding. Usually these oxides present a very small band gap since it opens up
via second or third-order perturbations. That is the situation in perovskite
(111) bilayers,\cite{lado} but here we see that the mechanism of gap opening in
principle could allow for larger gaps to appear. It is crucial to understand
what structures are prone to yield larger gaps, since these are required for
applications, but so far these have been quite elusive, at least using oxides
(with the exception of BaBiO$_3$,\cite{yan2013large} which is however a metal
if undoped). Being the band gap produced by SOC strength
mainly, our DFT calculations show that it is quite insensitive to the
exchange-correlation functional used, LDA or GGA yielding a very similar value
close to 18 meV. Comparing with our tight-binding model, we can then extract
the SOC strength by comparing the gap obtained with DFT and
that from the model (see Fig. \ref{model}d), leading to a $\lambda_{SOC}$ of
about 150-170 meV, which is a large value but consistent with the atomic number
of
Ag.

\subsection{Surface states}

The direct consequence of the non-triviality
of the band structure can be observed in finite geometries,
where
spin-polarized edge states show up. 
This can be easily checked with a tight binding model (in our case obtained
from the Wannierization described above) by building up
a finite system and calculating its electronic spectra.
In particular, in a semi-infinite
geometry two branches of edge states appear, which can
be calculated by solving the Dyson equation for each Bloch Hamiltonian
with wavevector parallel to the surface

\begin{equation}
G (k_x,E) = (E - h_0 (k_x) - t(k_x)^\dagger G (k_x,E) t(k_x))^{-1}
\label{sgreen}
\end{equation}
where $h_0(k_x)$ is the k-dependent intracell hopping matrix and
$t(k_x)$ the k-dependent intercell hopping matrix.
From the surface Green function (Eq.\ref{sgreen})
the spectral density can be obtained as

\begin{equation}
\rho(k_x,E) = -\frac{1}{\pi}\text{Tr}[\text{Im} [G(k_x,E+i0^+)]]
\end{equation}

\begin{figure}[t!]
\begin{center}
\includegraphics[width=\columnwidth]{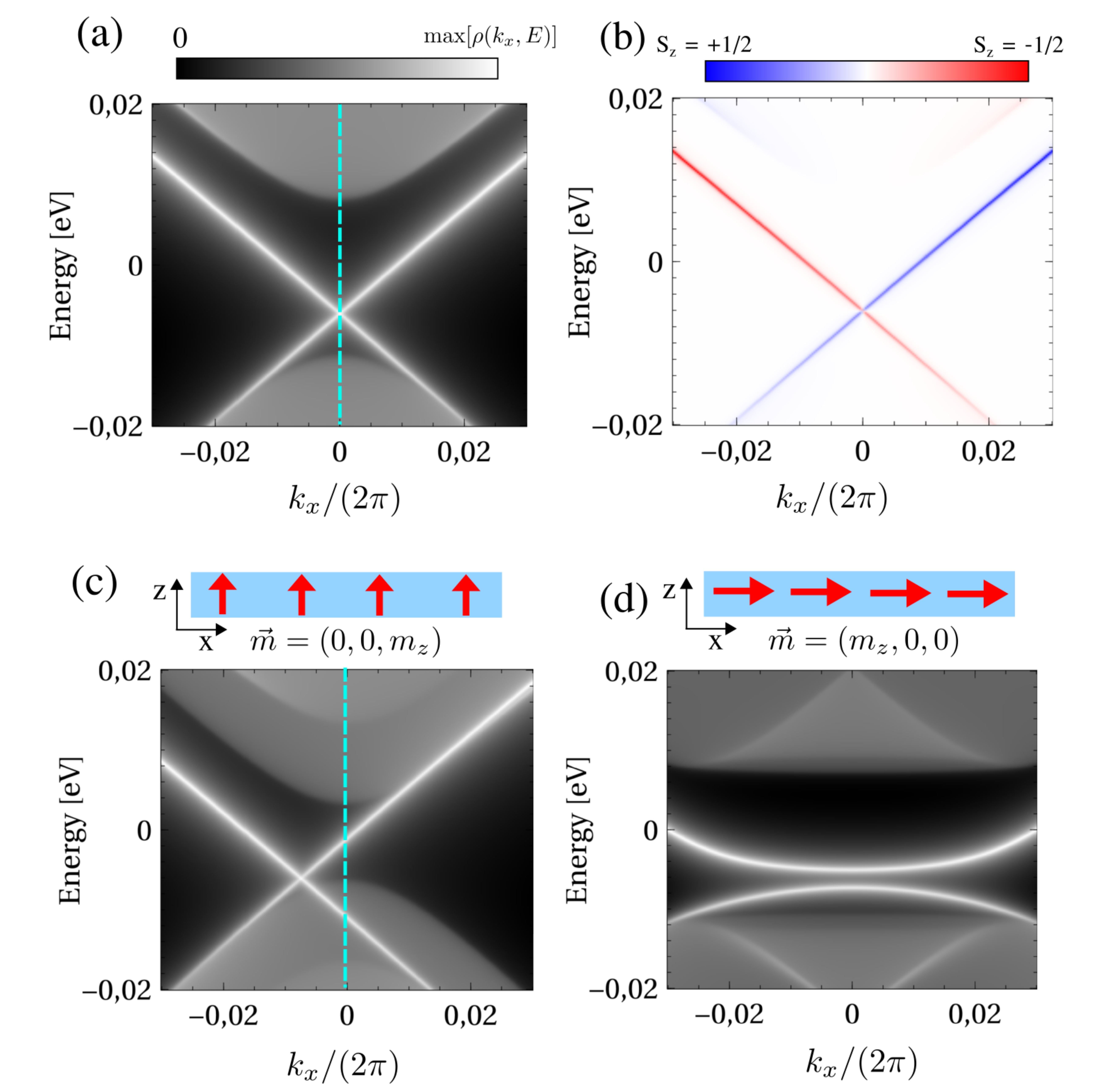}
\caption{(Color online.)
Surface spectral function (a) in armchair like edge, showing two branches of
gapless edge states plus the gapped bulk band structure. By calculating the spin character of those surface states (panel (b)), it is
observed that they show opposite spin polarization as expected from a quantum spin
Hall insulator due to spin-momentum locking.
When time reversal is broken by a uniform exchange coupling, the
conductance depends on the magnetization direction. For
off-plane exchange $m_z = 2$ meV the edge states remain gapless (c), whereas
for in-plane $m_x=10$ meV they acquire a small gap (d).
}\label{surf_states}
\end{center}
\end{figure}

The previous k-dependent spectral function (see Fig. \ref{surf_states}a)
shows the two surface branches edge states expected
for the semi-infinite geometry of a topological insulator.
This shows the gapless edge states that appear
well separated from the gapped bulk bands. Further insight on the topological
properties of those states
can be achieved by examining its spin polarization, which 
can be calculated from the spin spectral
function

\begin{equation}
\rho_z(k_x,E) = -\frac{1}{\pi}\text{Tr}[\text{Im} [S_z G(k_x,E+i0^+)]]
\end{equation}
where $S_z$ is the spin Pauli matrix.
This spin spectral function is shown in Fig. \ref{surf_states}b, and it confirms
the opposite spin polarization of the surface states that is produced by spin-momentum locking, typically obtained in QSHE systems like this.

The spin polarized states can be gapped out by perturbations
that break time reversal symmetry, such as local exchange fields.
That kind of perturbations can arise by interaction with a magnetic substrate,
substitutional magnetic impurities or an external magnetic field.
To understand the electronic properties in this
situation, we study what is the effect on
the surface edge states in the case that time reversal symmetry is broken. Under those circumstances,
the bulk $Z_2$ invariant cannot be defined anymore, but we can gain some
insight by studying
the surface states of the semi-infinite system when
adding a uniform local exchange to the Hamiltonian of the form:

\begin{equation}
H = H_0 + H_Z = H_0 + \sum_{i,s,s'}\vec m \cdot \vec S_{s',s} c_{i,s'}^\dagger c_{i,s}
\end{equation}
where $H_0$ is the Wannier Hamiltonian used previously.

We will consider two cases: when the magnetization is perpendicular to the
Ag hexagonal plane, and when it is parallel. For perpendicular exchange,
the nearly perfect spin polarization of the edge states allows them
to be resilient to a gap opening, still showing
gapless edge states (Fig. \ref{surf_states}c).
When the off-plane exchange field is large enough,
the bulk gap closes leading to a topological
metal, which has gapless edge states that coexist with normal
conduction electrons.
For in-plane exchange fields a gap opens up in the
edge channels\cite{rachel2014giant,lado2014magnetic} 
as shown in Fig. \ref{surf_states}d, although
the value of the gap opening is way smaller than the perturbation
in this armchair edge.
The bulk band structure remains gapped even when the in-plane exchange
field becomes
larger than the SOC gap. 
Something similar has been reported in oxides
before: VO$_2$/TiO$_2$ multilayers\cite{vpardo_Dirac,sD_banerjee} have been
shown to be Chern insulators\cite{vo2_tio2_vanderbilt} when magnetization is
along the off-plane direction and open a trivial gap when the magnetization
lies inside the plane.\cite{vo2_tio2_mit}

\begin{figure}[t!]
\begin{center}
\includegraphics[width=\columnwidth]{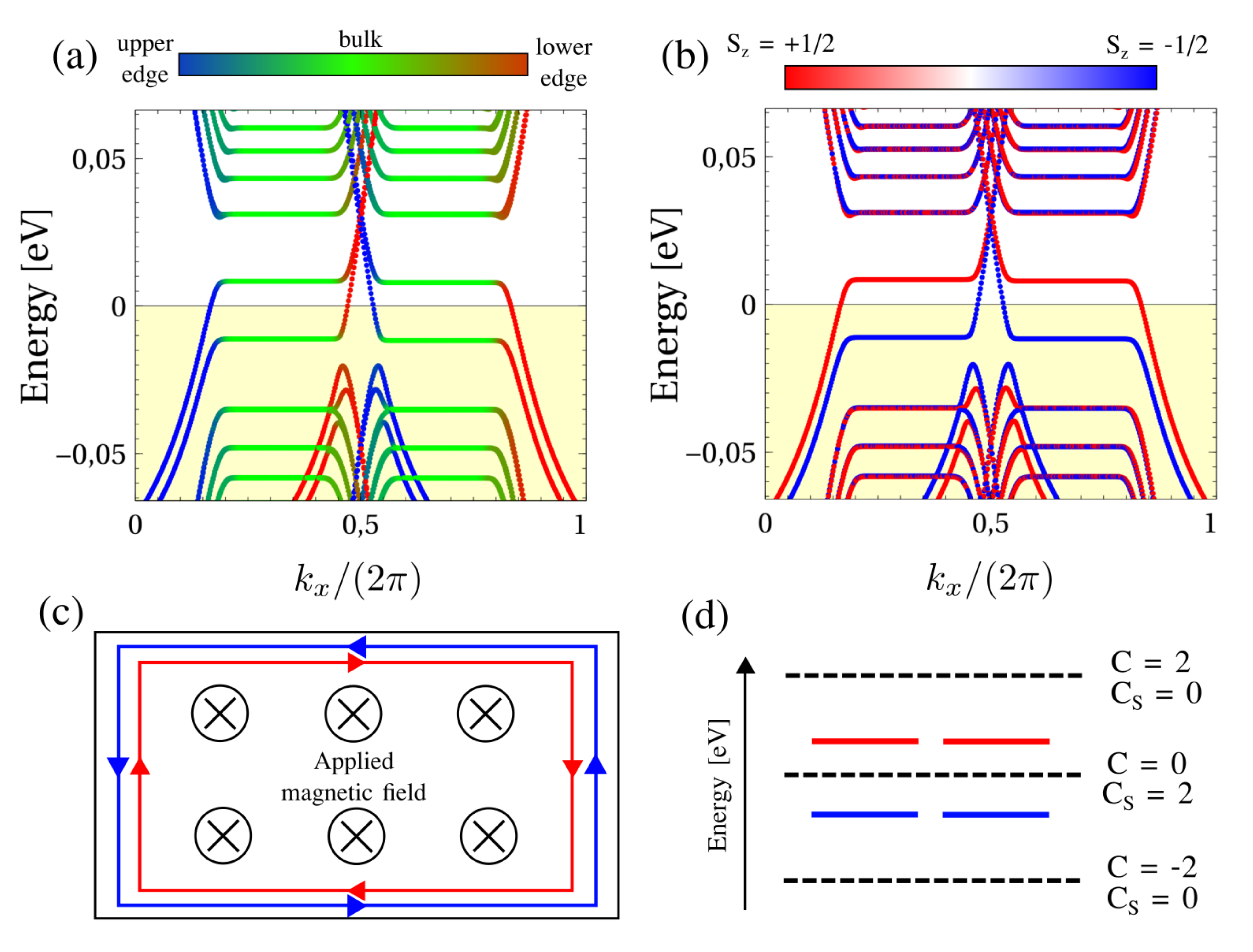}
\caption{(Color online.) (a) Band structure
of a quantum Hall slab of the Ag-Zn compound, showing the position (a) and
spin flavor (b) of the different eigenstates. The quantum spin Hall states
survive even when time reversal symmetry is broken by a gauge magnetic field,
and turns the system a ferromagnet even without interactions. Panel (c) shows a
sketch of the quantum spin Hall configuration and (d) shows the different
insulating states as function of the zero Landau level
filling, quantum spin Hall at filling 0, and
quantum Hall for filling $\pm 2$.
}
\label{fig:hall}
\end{center}
\end{figure}

\subsection{Quantum Hall effect}

The Dirac-like low-energy spectra suggests that the system will
show an unconventional Landau level (LL) spectrum very much like graphene.
In the absence of SOC, the Dirac dispersion will yield
a set of 4 zero Landau levels: one per spin and valley.
The topological gap is equivalent to a mass in the K and K' points of the Brillouin zone
that follows $m = s_z \kappa $, where $s_z$ labels spin and
$\kappa$ labels valley. When a magnetic field is turned on,
the zero Landau level will develop a splitting following
$\Delta = m \kappa = s_z$, therefore independent on
the valley. The previous splitting is equivalent to the one
obtained in a Dirac ferromagnet\cite{young2012spin,fertig2006luttinger} 
but with
the zero LL spin splitting coming from SOC instead of Zeeman, 
and automatically realizes
the quantum spin Hall effect.
Using the tight binding model derived before, we build a quantum Hall
bar, where the magnetic field is included by means
of the Peierls substitution $t_{ij} \rightarrow t_{ij}e^{i\phi_{ij}}$,
with $\phi_{ij} = \frac{eB}{\hbar}(x_i -x_j)(y_i+y_j)$ and $x_i,y_j$
the positions of the Wannier charge centers. The Zeeman
term of the magnetic field is neglected provided
that the splittings created by SOC are larger. The band structure obtained
is shown in Fig. \ref{fig:hall}a,b, where the position and spin
are represented by the color of the eigenvalue. 

When the system is with the
$d_{z^2}$ manifold half filled, the spin-down zero LL are filled, whereas
the spin-up are empty. In this situation,
the Hall bar is the quantum spin Hall state, characterized
by a spin Chern number $C_S=C_{\uparrow}-C_{\downarrow} = 2$, and
two counter-propagating edge channels (Fig \ref{fig:hall}c). It
is interesting to note that this quantum spin Hall state
is not protected by time reversal symmetry but by conservation of
$S_z$, and that the total Chern number is $C=0$, so that the
system will be vulnerable to perturbations
creating spin mixing. In addition, this QSH state
is adiabatically connected to the one occurring in the absence of SOC.
When the Fermi energy is moved away from half filling (Fig. \ref{fig:hall}d) of the
$d_{z^2}$ states and all
the zero Landau levels are filled, the system no longer realizes the
QSH state. Instead, it enters into a normal quantum Hall state
with $C=2$ and two co-propagating edge states.

\section{Concluding remarks.}

To summarize, in this paper we present ab initio calculations on 
the oxide 
InZn$_{1/3}$Ag$_{1/3}$V$_{1/3}$O$_3$
based on the structure of the recently synthesized
InCu$_{2/3}$V$_{1/3}$O$_3$.
We have emphasized the Dirac-like low-energy
electronic structure that occurs when magnetism is switched off and
two-dimensionality is enforced.
The low-energy properties, on both sides of the Fermi level, are shown to be
dominated by Ag
$d_{z^2}$ orbitals.
The unexpectedly simple low-energy effective model compares with the rather
complex crystal structure and deep energy electronic mixing.
In the same fashion of other Dirac materials, the inclusion of spin-orbit coupling leads to a topological insulating state, with a band
gap scaling linear with SOC strength. The non-triviality
of the gap was confirmed by the calculation of the $Z_2$ invariant
and edge states spectral functions. The study of this system provides some
clues on how one can design topologicality using oxides, and in particular to
understand how the topological gap can be tuned or even enhanced. We have
compared the results obtained for this system with other oxide structures built
from a honeycomb lattice, which can be even qualitatively very different. We
have also analyzed the material using a tight-binding model obtained through a Wannierization procedure that fits perfectly the DFT 
bands around the Fermi level. Our calculations using the tight-binding
Hamiltonian describe the behavior of such system in a quantum Hall experiment,
and show that its properties would be very similar to the response of graphene,
another single-band system (of p$_z$ symmetry in that case) with honeycomb
structure. Thus, this system we propose would be a close d-electron analogue of
graphene.

\acknowledgments

This work was supported by Xunta de Galicia under the Emerxentes Program via the
project no. EM2013/037 and the MINECO via project MAT2013-44673-R. V.P.
acknowledges support from the MINECO of Spain via the Ramon y Cajal program
RyC-2011-09024. J. L. Lado acknowledges financial support
by Marie-Curie-ITN Grant No. 607904-SPINOGRAPH.

\section{Computational procedures}

Ab initio electronic structure calculations based on the density functional
theory (DFT)\cite{hk,ks} have been performed
using two all-electron full
potential codes ({\sc wien2k}\cite{wien} and {\sc
Elk}\cite{elk}) and the pseudopotential-based
Quantum Espresso code.\cite{giannozzi2009quantum}
Unless stated otherwise, structural optimizations and band structure
calculations were performed with GGA-PBE\cite{gga}.

 {\sc wien2k} calculations
were performed with a converged k-mesh, a value of
R$_{mt}$K$_{max}$= 7.0, and
spin-orbit coupling was introduced
in a second variational manner using the scalar
relativistic approximation.\cite{singh}
Elk calculations were carried out
with the non-collinear formalism and spin orbit coupling
in order to calculate the
topological gap as well as to check that the Ag based compound is
non magnetic.
Quantum Espresso calculations were carried out using PAW
pseudopotentials.\cite{kresse1999ultrasoft} 
The structures were relaxed and the different approaches gave analogous results.

The Wannierization\cite{mostofi2008wannier90,marzari1997maximally,souza2001maximally,marzari2012maximally}
 is performed in the non relativistic Quantum Espresso
PAW calculation. The frozen window is chosen
in the interval [-0.9,0.6] eV so that it contains the
low energy dz$^2$ bands, whereas the outer window is chosen
in the interval [-7,0.6] eV around the Fermi energy. The projections
used are the d-manifold of the Ag atoms, giving rise to a $10\times10$
tight binding Hamiltonian. Spin orbit coupling is included
afterwards as an $\vec L \cdot \vec S$ term in the
Wannier tight binding model obtained, turning the Hamiltonian into a
$20\times20$ matrix.

\bibliography{honeycomb}

\end{document}